\documentclass{lipics-v2021}

\usepackage{cite}
\usepackage{amsmath,amssymb,amsfonts,amsthm}
\usepackage{algorithmic}
\usepackage{graphicx}
\usepackage{textcomp}
\usepackage{url}
\usepackage{orcidlink}
\usepackage{xcolor}
\usepackage{stmaryrd}
\usepackage{mathtools}
\usepackage{tikz}
\usepackage{booktabs}
\usetikzlibrary{positioning}

\makeatletter
\DeclareFontEncoding{LS1}{}{}
\DeclareFontEncoding{LS2}{}{\noaccents@}
\makeatother

\DeclareFontSubstitution{LS1}{stix}{m}{n}
\DeclareFontSubstitution{LS2}{stix}{m}{n}

\DeclareSymbolFont{symbols2}{LS1}{stixfrak}{m}{n}
\DeclareSymbolFont{arrows3}{LS2}{stixtt}{m}{n}
\DeclareSymbolFont{symbols3}{LS1}{stixbb}{m}{n}

\DeclareSymbolFontAlphabet{\mathbb}{symbols3}

\nolinenumbers
\hideLIPIcs
\title{Short proofs without interference}
\author{Adrian Rebola-Pardo}%
    {Vienna University of Technology \and Johannes Kepler University}%
    {adrian.rebola@tuwien.ac.at}%
    {0000-0001-9234-4377}{}
\authorrunning{A.\ Rebola-Pardo}
\Copyright{Adri\'an Rebola-Pardo}

\DeclareMathSymbol{\smashtimes}{\mathbin}{arrows3}{"A4}
\DeclareMathSymbol{\vzigzag}{\mathord}{symbols2}{"39}

\DeclarePairedDelimiter\Xpar{(}{)}
\DeclarePairedDelimiter\Xbra{\{}{\}}

\DeclarePairedDelimiter\Xang{\langle}{\rangle}
\DeclarePairedDelimiterX\Xset[2]{\{}{\}}{#1 : #2}
\DeclarePairedDelimiterX\XparBar[2]{(}{)}{#1\,\delimsize\vert\,\mathopen{}#2}
\DeclarePairedDelimiterX\XparBarBar[3]{(}{)}{#1\,\delimsize\vert\,\mathopen{}#2\,\delimsize\vert\,\mathopen{}#3}

\newcommand{\subst}[2]{\left.#1\right|_{#2}}
\newcommand{\conflict}[1]{#1 \vdash \mathord{\smashtimes}}
\newcommand{\pproof}[3]{#1\colon #2 \vdash #3}
\newcommand{\rup}[1]{\Xpar{\textbf{\upshape a:}\;#1}}
\newcommand{\del}[1]{\Xpar{\textbf{\upshape d:}\;#1}}
\newcommand{\sr}[2]{\Xpar{\textbf{\upshape r:}\;#1,#2}}
\newcommand{\wsr}[3]{\Xpar{\textbf{\upshape w:}\;#1,#2,#3}}

\newcommand{\test}{\mathord{?}}
\newcommand{\emptylist}{\mathord{\vzigzag}}

\newcommand{\branch}[2]{\nabla\Xpar{#1\colon\,#2}}
\newcommand{\acc}[1]{\mathrm{acc}\Xpar{#1}}
\newcommand{\ctx}[1]{\mathrm{ctx}\Xpar{#1}}
\newcommand{\at}{\mathbin{@}}
\newcommand{\static}{\mathord{\downarrow}}
\newcommand{\wqed}{\textbf{\upshape qed}}
\newcommand{\wlem}{\textbf{\upshape lem}}
\newcommand{\wctx}{\textbf{\upshape ctx}}
\newcommand{\welim}{\textbf{\upshape elim}}
\newcommand{\iqed}{\wqed}
\newcommand{\ilem}[2]{\Xpar{\wlem\;#1,#2}}
\newcommand{\ictx}[2]{\Xpar{\wctx\;#1,#2}}
\newcommand{\ielim}{\welim}
\newcommand{\wsrtrans}[2]{\textrm{dyn}(#1, #2)}

\newcommand{\dynimp}{\mathrel{\vdash_{\mathrm{dyn}}}}

\begin{document}
\maketitle

\begin{abstract}
Interference is a phenomenon on proof systems for SAT solving
that is both counter-intuitive and bothersome when developing
proof-logging techniques.
However, all existing proof systems that can produce short proofs
for all inprocessing techniques deployed by SAT present this feature.
Based on insights from propositional dynamic logic,
we propose a framework that eliminates interference while
preserving the same expressive power of interference-based proofs.
Furthermore, we propose a first building blocks towards
RUP-like decision procedures for our dynamic logic-based frameworks,
which are essential to developing effective proof checking methods.
\end{abstract}

\section{Introduction}

A plethora of proof systems for SAT solving has been developed
over the last decade, with DRAT~\cite{Heule16} becoming a \emph{de facto} standard.
Limitations of DRAT have been iteratively tackled
by both extensions and alternatives,
with more expressive redundance rules~\cite{HeuleKB17,BussT19},
support for unit deletions~\cite{AltmanningerP20},
incremental solving-friendly proofs~\cite{FazekasPFB24},
handling of pseudo-Boolean constraints~\cite{GochtN21},
methods for simultaneous symmetry breaking~\cite{BogaertsGMN23},
redundance lemmas, and fixpoint-free trimming~\cite{Rebola-Pardo23}.

All these proof systems share a feature we call the
\emph{accumulated formula}: a formula is carried over
throughout the proof, initially set to the formula being refuted
and modified by each proof step by adding or deleting a single constraint;
proof rules ensure that the accumulated formula at each point
preserves the satisfiability status of its predecessor.
They share as well a phenomenon known as \emph{interference}~\cite{HeuleK17B},
where inference rules are non-monotonic
and depend on the whole accumulated formula.
In other words, if $C$ can be added to the accumulated formula $F$
by some proof step, it is not true in general that
$C$ can be added to $F \cup G$ by the same proof step.
Interference inadvertently creates a myriad
of counter-intuitive situations~\cite{JarvisaloHB12,HeuleB15b,Rebola-PardoB18,Rebola-Pardo23}:
\begin{itemize}
\item Deleting a (correctly derived) constraint from the formula
may turn an incorrect proof correct.
\item Constraints derived by two different proofs from the same formula
may be mutually inconsistent.
\item Trimming (i.e.\ removing parts of a proof not needed to derive
unsatisfiability) may render a proof incorrect.
\item Temporarily deriving a constraint as a lemma to prove another constraint
may render the latter underivable.
\end{itemize}

Given the hurdles brought about by interference,
it is not surprising that good reasons exist for the popularity of
interference-based proof systems.
All these proof systems bypass intrinsic limitations of resolution proofs.
Some techniques deployed by SAT solvers,
including Gaussian elimination~\cite{SoosNC09}, symmetry breaking~\cite{AloulRMS03} and
bounded variable addition~\cite{MantheyHB12} must yield resolution proofs
that are worst-case exponentially larger than their own execution~\cite{Haken85,Urquhart87,Urquhart99}.
Interference-based proofs greatly alleviate this problem,
since no exponential lower bounds for these proof systems are known~\cite{KieslRH18,HeuleB18}.

In this work we argue that we must not choose between non-interference and
short proofs. We present an interference-free proof system with proofs
as short as interference-based systems.
To achieve this, constraints are annotated with a \emph{program},
encoding the idea that a property holds after the execution of that program.
This idea is inspired by \emph{propositional dynamic logic (PDL)},
a multimodal logic where each program defines a modality~\cite{FischerL79}.
Our framework restricts PDL to a fragment amenable to
proofs for SAT solving. Finally, we present a translation from DSR/WSR proofs
into our proof system, and we showcase how non-interference can be used
in our system for proof composition.

\section{Preliminaries}
\label{sec:prelim}

Much of this paper deals with lists,
which we denote by writing their items juxtaposed;
the empty list is denoted $\emptylist$,
and concatenation is expressed by juxtaposition as well.

We consider a nonempty set of \emph{variables}.
A \emph{(total) assignment} $I$ maps each variable to either $0$ or $1$.
A \emph{logical expression} is any expression $e$ over which we define
a semantics, i.e.\ conditions under which an assignment $I$ \emph{satisfies} $e$;
we write this as $I \vDash e$. We call a logical expression \emph{satisfiable}
if there is some assignment $I$ with $I \vDash e$, and \emph{unsatisfiable} otherwise.
Given two logical expressions $e_1, e_2$, we say that $e_1$ \emph{entails} $e_2$
if $I \vDash e_1$ implies $I \vDash e_2$ for all assignments $I$;
we write this as $e_1 \vDash e_2$. If both $e_1 \vDash e_2$ and $e_2 \vDash e_1$ hold,
then we say they are \emph{equivalent}, and we write $e_1 \equiv e_2$.
We will also consider \emph{logical relations}, which are expressions $r$
over which we define a relational semantics, i.e.\ conditions under which
an assignment $I$ \emph{transitions} into another assignment $J$ through $t$;
we write this as $I \otimes J \vDash t$.
We say that $t$ \emph{progresses} over $I$ whenever
there exists some assignment $J$ with $I \otimes J \vDash t$.

A \emph{literal} is a logical expression which is
either a variable $x$ (with $I \vDash x$ iff $I(x) = 1$)
or its negation $\neg x$ (with $I \vDash \neg x$ iff $I(x) = 0$).
For simplicity of exposition, we consider the expressions $\top$ and $\bot$ as literals too,
with $I \vDash \top$ and $I \not\vDash \bot$ for all assignments $I$.
The \emph{complement} $\overline{l}$ of a literal $l$ is:
\begin{align*}
\overline{x} = {} & \neg x & \overline{\neg x} = {} & x &
\overline{\top} = {} & \bot & \overline{\bot} = {} & \top
\end{align*}
An \emph{(atomic) substitution} $\sigma$ maps each literal to a literal while fulfilling
$\sigma(\top) = \top$ and $\sigma(\overline{l}) = \overline{\sigma(l)}$ for all literals $l$.
Since we are not assuming a finite set of variables,
some substitutions might be hard or impossible to record in a file.
In these cases we will require $\sigma$ to be \emph{finite},
i.e.\ $\sigma(x) = x$ holds for all but a finite number of variables $x$.
Finite substitutions can be recorded by simply writing its non-identical mappings.
% A finite substitution with $\sigma(x_i) = l_i$ for $1 \leq i \leq n$
% and $\sigma(x) = x$ for $x \notin \Xbra{x_1,\dots,x_n}$ can then be recorded
% as $\Xbra{x_1 \mapsto l_1, \dots, x_n \mapsto l_i}$.

Given an assignment $I$ and a substitution $\sigma$,
we define the assignment $I \circ \sigma$ where
$(I \circ \sigma)(x) = 1$ iff $I \vDash \sigma(x)$ for all variables $x$.
We also define the \emph{composition} $\tau \circ \sigma$ of two substitutions,
which is a substitution with $(\tau \circ \sigma)(l) = \tau(\sigma(l))$.
\begin{lemma}
$I \circ (\tau \circ \sigma) = (I \circ \tau) \circ \sigma$ for any assignment $I$.
Furthermore, if $\tau$ and $\sigma$ are finite, then so is $\tau \circ \sigma$.
\end{lemma}
% Given variables $x_1,\dots,x_n$ and literals $l_1,\dots,l_n$,
% we denote by $\Xbra{x_1 \mapsto l_1,\dots,x_n \mapsto l_n}$ the substitution
% with $\sigma(x_i) = l_i$ for $1\leq i \leq n$ and $\sigma(x) = x$
% for all variables $x \notin \Xbra{x_1,\dots,x_n}$.

\subsection{Constraints and plain proofs}
\label{ssc:prelim:constraints}

Typically, proof systems for SAT solving only handle logical expressions of
a very restricted family, most often clauses or pseudo-Boolean constraints.
The framework we present is very general on the choice of such expressions,
which we call \emph{constraints}.
A \emph{formula} $F$ is a finite set of constraints;
formulas are themselves logical expressions,
with $I \vDash F$ if $I \vDash C$ for all $C \in F$.
Our framework requires the following from the family of logical expressions
we admit as constraints:
\begin{enumerate}
    \item\label{itm:cstreqs:lit}
    For any literal $l$ we can construct a constraint $C$ with $C \equiv l$;
    we simply denote this constraint as $l$.
    \item\label{itm:cstreqs:neg}
    We can construct a constraint $\overline{C}$, called the \emph{negation} of $C$,
    such that $I \vDash \overline{C}$ iff $I \not\vDash C$.
    \item\label{itm:cstreqs:infer}
    There is a sufficient, computable criterion called \emph{conflict detection}
    for $F$ to be unsatisfiable; if this criterion is satisfied we write
    $\conflict{F}$.
    We further require that:
    \begin{itemize}
    \item $\conflict{F}$ is complete criterion for unsatisfiability
    when $\Xbra{C, \overline{C}} \subseteq F$
    for any constraint $C$, and when $\bot \in F$.
    \item If $F \subseteq F^\prime$ and $\conflict{F}$, then $\conflict{F^\prime}$.
    \end{itemize}
    \item\label{itm:cstreqs:subst}
    For each substitution $\sigma$, we can construct a constraint $\subst{C}{\sigma}$,
    called the \emph{reduct} of $C$ by $\sigma$, such that $I \vDash \subst{C}{\sigma}$
    iff $I \circ \sigma \vDash C$.
    We also define reducts of formulas as
    $\subst{F}{\sigma} = \Xset{\subst{C}{\sigma}}{C \in F}$.
\end{enumerate}

Note that these are minimum requirements for our framework to be well-defined.
For the framework to be \emph{useful},
\eqref{itm:cstreqs:lit}, \eqref{itm:cstreqs:neg} and \eqref{itm:cstreqs:subst}
must be extremely efficient, and their outputs not be significantly larger than $C$;
and conflict detection in \eqref{itm:cstreqs:infer} must be \emph{quite} efficient to compute,
and moderately powerful.
Because parts of our presentation involve DRAT-like proofs,
we explain how clauses meet these requirements.
Note, however, these are not quite far-fetched:
pseudo-Boolean constraints~\cite{RousselM21}, and parity constraints of the form
$l_1 \oplus \dots \oplus l_n$~\cite{LaitinenJN12}, fulfill them too.

Let $l_1,\dots,l_n$ be literals.
A \emph{clause} $C$ is an expression $l_1 \vee \dots \vee l_n$
with $I \vDash C$ iff $I \vDash l_i$ for some $1 \leq i \leq n$;
a \emph{cube} $Q$ is  an expression $l_1 \wedge \dots \wedge l_n$
with $I \vDash Q$ iff $I \vDash l_i$ for all $1 \leq i \leq n$.
Negations and reducts are defined by:
\begin{align*}
\overline{l_1 \vee \dots \vee l_n} = {} & \overline{l_1} \wedge \dots \wedge \overline{l_1} \\
\overline{l_1 \wedge \dots \wedge l_n} = {} & \overline{l_1} \vee \dots \vee \overline{l_1} \\
\subst{(l_1 \vee \dots \vee l_n)}{\sigma} = {} & \sigma(l_1) \vee \dots \vee \sigma(l_n) \\
\subst{(l_1 \wedge \dots \wedge l_n)}{\sigma} = {} & \sigma(l_1) \wedge \dots \wedge \sigma(l_n)
\end{align*}
We call formulas of this family of constraints
\emph{conjunctive normal form (CNF)} formulas\footnote{CNF formulas are often
presented as only containing clauses. Treating both clauses and clauses
as basic constraints makes this presentation much simpler,
and it creates no significant issues.}.
Conflict detection is given by
$\conflict{F}$ iff \emph{unit propagation (UP)}~\cite{DavisP60} over $F$ reports a conflict.
Intuitively, UP (also known as \emph{Boolean constraint propagation})
builds a set of literals implied by individual clauses
together with previously implied literals,
reporting a conflict when complementary literals are implied;
we refer the interested reader to~\cite{SilvaLM21} for more details.
It is easy to check that unit propagation satisfies the requirements in
\eqref{itm:cstreqs:infer}.

\subsection{Interference-based proofs}
\label{ssc:prelim:proofs}

In this paper we focus on the DSR and WSR proof systems,
since many other interference-based proof systems are
essentially captured by these (with the notable exception of VeriPB proofs with
dominance-based redundancy~\cite{BogaertsGMN23}).

Both \emph{Deletion Substitution Redudancy (DSR)}~\cite{BussT19} and
\emph{Weak Substitution Redundancy (WSR)}~\cite{Rebola-Pardo23} are clause-based,
with proofs expressed as a list of \emph{instructions}.
These instructions are based on \emph{redundancy rules}:
\begin{itemize}
\item A clause $C$ is a \emph{reverse unit propagation (RUP)}~\cite{GoldbergN03} clause over
a CNF formula $F$ if $\conflict{F \cup \Xbra{\overline{C}}}$.
\item $C$ is a \emph{substitution-redundant (SR)}~\cite{BussT19} clause over $F$ upon
a substitution $\sigma$ if $\conflict{\Xbra{\overline{\subst{C}{\sigma}}}}$
and for all $D \in F$ we have $\conflict{F \cup \Xbra{\overline{C}, \overline{\subst{D}{\sigma}}}}$.
\item $C$ is a \emph{weak substitution-redundant (WSR)}~\cite{Rebola-Pardo23,GochtN21} clause over $F$
upon $\sigma$ modulo a CNF formula $G$ if for all $D \in F \setminus G \cup \Xbra{C}$ we have
$\conflict{F \cup \Xbra{\overline{C}, \overline{\subst{D}{\sigma}}}}$.
\end{itemize}
DSR allows instructions $\del{C}$, $\rup{C}$ and $\sr{C}{\sigma}$,
and WSR allows only the instruction $\wsr{C}{\sigma}{G}$,
for clauses $C$, finite substitutions $\sigma$ and CNF formulas $G$.
Given a list of instructions $\pi$ and a CNF formula $F$,
the \emph{accumulated formula} by $\pi$ over $F$,
denoted $\acc{F, \pi}$, is defined by:
\begin{align*}
    \acc{F, \emptylist} = {} & F \\
    \acc{F, \pi\,\del{C}} = {} & F \setminus \Xbra{C} \\
    \acc{F, \pi\,\rup{C}} = {} & \acc{F, \pi} \cup \Xbra{C} \\
    \acc{F, \pi\,\sr{C}{\sigma}} = {} & \acc{F, \pi} \cup \Xbra{C} \\
    \acc{F, \pi\,\wsr{C}{\sigma}{G}} = {} & \acc{F, \pi} \setminus G \cup \Xbra{C}
\end{align*}
\emph{Derivations} from a CNF formula $F$ are defined recursively as follows:
\begin{itemize}
\item $\emptylist$ is a derivation from $F$.
\item $\pi\,\del{C}$ is a derivation from $F$ if so is $\pi$.
\item $\pi\,\rup{C}$ is a derivation from $F$ if $\pi$ is a derivation from $F$
and $C$ is a RUP clause over $\acc{F, \pi}$.
\item $\pi\,\sr{C}{\sigma}$ is a derivation from $F$ if $\pi$ is a derivation from $F$
and $C$ is an SR clause over $\acc{F, \pi}$ upon $\sigma$.
\item $\pi\,\wsr{C}{\sigma}{G}$ is a derivation from $F$ if $\pi$ is a derivation from $F$
and $C$ is an WSR clause over $\acc{F, \pi}$ upon $\sigma$ modulo $G$.
\end{itemize}
A derivation $\pi$ from $F$ is called a \emph{refutation} of $F$ if
$\conflict{\acc{F, \pi}}$.
For a derivation $\pi$ from $F$, it is not in general true that $F \vDash \acc{F, \pi}$.
This is because SR and WSR clauses are not necessarily entailed by $F$,
though RUP clauses are~\cite{JarvisaloHB12,Rebola-PardoS18,Rebola-Pardo23}.
However, it can be shown that if $F$ is satisfiable, then so is $\acc{F, \pi}$
(in particular, if $\pi$ is a refutation, then $F$ is unsatisfiable).
This is shown by an inductive argument over the following result,
whose various parts are shown in~\cite{GoldbergN02,HeuleHW13,BussT19,Rebola-Pardo23}.

\begin{theorem}
\label{thm:dsr-semantics}
Let $I \vDash F$, and consider a single-instruction derivation $\pi$.
\begin{itemize}
\item If $\pi = \del{C}$ or $\pi = \rup{C}$, then $I \vDash \acc{F, \pi}$.
\item If $\pi = \sr{C}{\sigma}$ or $\pi = \wsr{C}{\sigma}{G}$,
then let $J$ be $I$ if $I \vDash C$, and $I \circ \sigma$ otherwise.
Then, $J \vDash \acc{F, \pi}$.
\end{itemize}
\end{theorem}

As pointed out in~\cite{Rebola-PardoS18,Rebola-Pardo23},
interference-based rules, including RAT~\cite{JarvisaloHB12},
PR~\cite{HeuleKB17}, SR~\cite{BussT19,GochtN21} and WSR~\cite{Rebola-Pardo23},
perform reasoning without loss of generality.
Deriving $\sr{C}{\sigma}$ from $F$ can be understood as follows:
under $F$, we can assume $C$ is satisfied; if this is not the case,
then $\sigma$ transforms the current assignment into an assignment that satisfies
both $F$ and $\sigma$.
Hence, that satisfiability is preserved by the derivation is an understatement:
Theorem~\ref{thm:dsr-semantics} tells us how to read from a derivation
a procedure that transforms any assignment satisfying $F$ into
an assignment satisfying $\acc{F, \pi}$.

This view of SR as reasoning without loss of generality
points to the defining feature of interference.
In order to introduce the assumption without loss of generality,
$F$ must be satisfied after applying the transformation.
This means that, unlike other reasoning rules such as resolution
or RUP, strengthening $F$ by introducing new clauses
may invalidate that reasoning step.
This non-monotonic behavior has widespread consequences.
For one, proofs are not tree-shaped as usual in proof systems,
since these reasoning steps depend on the whole formula.
This creates problems for trimming~\cite{Rebola-Pardo23},
requiring even more complex rules just to be able to
remove parts of the proof
\emph{that are not involved in deriving a contradiction}.
For another, proofs lose compositionality,
so concatenating proofs does not yield a proof,
which is a desirable trait when generating proofs
in a distributed or incremental SAT solver.

\subsection{Interference as a hurdle}
\label{ssc:prelim:hurdle}

Despite much publicity as a positive feature enabling
short proofs and novel reasoning techniques~\cite{HeuleK17B},
interference can also \emph{prevent} reasoning that
is almost trivial without interference.
This becomes apparent when trying to compose
interference-based proofs from smaller proofs.

Let us consider unsatisfiable CNF formulas $F_1,\dots,F_n$.
Choose distinct fresh variables $z_1,\dots,z_n$,
and define
\begin{align*}
C = {} & \neg z_1 \vee \dots \neg z_n \\
G_i = {} & \Xset{C \vee z_i}{C \in F_i} \\
G = {} & G_1 \cup \dots \cup G_n \cup \Xbra{C}
\end{align*}
for $1 \leq i \leq n$.
Given resolution refutations of $F_i$
(obtained by e.g.\ a purely CDCL SAT solver),
we can turn them into derivations of the clause $z_i$ from $G_i$
by appending $z_i$ in each derivation clause.
Concatenating these derivations constructs a derivation of
the formula $H = \Xbra{z_1, \dots, z_n, C}$ from $G$,
and extending this to a
refutation of $G$ is straightforward.

This seemingly innocent reasoning becomes much harder
as soon as the SR and WSR rules are involved
in solving $F_i$.
Let us assume that each $F_i$ has a distinct
order-2 symmetry $\sigma_i = \Xbra{x_i \mapsto y_i, y_i \mapsto x_i}$
for variables $x_i, y_i$, that is, $\subst{F_i}{\sigma_i} = F_i$.
For the sake of the argument,
we assume the $F_i$ have the same sets of variables
and syntactically different from each other in a non-trivial way;
while we are at it, let us make $F_i$ \emph{really hard} to solve.

If a formula has a symmetry, introducing a \emph{symmetry breaking predicate}
can have dramatic effects over solving runtime~\cite{AloulRMS03,AndersBR24,Urquhart99}.
In our particular case, the symmetry breakers for the $F_i$
are $B_i = \neg x_i \vee y_i$; each of these is an SR clause over $F_i$
upon the substitution $\tau_i = \Xbra{x_i \mapsto \bot, y_i \mapsto \bot}$~\cite{HeuleHW15}.
Note that $B_i$ is not implied by $F_i$, but inserting $B_i$ does not change
its satisfiability status.
Then, one possible way to solve $G$ goes as follows.
For each of the $G_i$, run some symmetry detection tool to detect $\sigma_i$.
Then run a CDCL solver on $G_i$ until the clause $z_i$ is derived.
Since the variable $z_i$ was not involved in the symmetry,
it is actually implied by $G_i$, even if $B_i$ was used to derive it.
Then, we have derived again $H$, which is easy to refute.

We can express the derivation of $z_i$ from $G_i$ through a
DSR derivation $\pi_i$ given by $\sr{B_i}{\tau_i}$ followed by
$\rup{D}$ instructions for each clause $D$ derived by the CDCL solver~\cite{GoldbergN03}.
Unfortunately, there is no reasonable way to combine the $\pi_i$
to derive $H$, even if this is perfectly sound.
The culprit is interference: the DSR and WSR rules depend on the whole formula.
Since $\sigma_i$ is only a symmetry of $G_i$
and cannot be extended to a symmetry of $G$
(because we required the $F_i$ to have the same variables),
the instruction $\sr{B_i}{\tau_i}$ fails during proof checking.

This example showcases the problems that arise when using
interference-based reasoning, in particular proof compositionality;
for example, \cite{HeuleB15b} only allows RUP steps in proofs
when performing parallel composition.
Similar problems arise in incremental solving,
where the notion of \emph{clean clauses} also exploits that
substitutions do not affect some clauses~\cite{FazekasBS19,FazekasPFB24}.
On the other hand, interference-based proofs
are able to handle much more complex reasoning,
even when eventually only implied clauses remain in the formula~\cite{PhilippR16,GochtN21}.

The rest of this paper is devoted to showing that
this ability has \emph{nothing at all} to do with interference.
Rather, interference is caused by our proof systems forcing us
to operate on an accumulated formula.
We show that building similarly capable, SAT-friendly proof systems
without interference is possible,
if one is willing to take a detour through modal logic.

\section{A SAT-friendly dynamic logic}
\label{sec:dynlog}

In this section we present a version of \emph{dynamic propositional logic (PDL)},
a multimodal logic developed during the 1980s to reason about programs~\cite{FischerL79}.
PDL assigns to each program a modality, with worlds representing program states.
A state $s^\prime$ is accessible from another state $s$ through
the modality of a program $\varepsilon$ if executing $\varepsilon$ in
a system in state $s$ may leave the system in state $s^\prime$.
In this setting, the necessity operator for $\varepsilon$
constructs safety conditions~\cite{ClarkeES86}, stating that a formula holds
after executing $\varepsilon$; the possibility operator
corresponds then to liveness conditions.

This modal view of programs is by no means unique to PDL.
That programs themselves are specified within the logic is, though.
PDL was designed to reason about \emph{general} properties of programs,
so its most basic program constructs are program constants,
which are arbitrarily interpreted. Valid formulas in PDL, thus,
must be true for all programs. The program language is enriched
with test, choice and composition operators;
adding a Kleene star operator allows the expression of loops.

The connection between PDL and interference-based proofs is uncanny.
Theorem~\ref{thm:dsr-semantics} shows that,
when we are deriving a formula through a DSR/WSR derivation,
what we are \emph{really} doing is inferring properties
that hold after applying a transformation.
This transformation, moreover, has a very clear computational content:
if we see assignments as a memory bitfield
(where each variable represents the address of a single bit),
the transformation defined in Theorem~\ref{thm:dsr-semantics}
checks if the current memory state satisfies the condition $\overline{C}$,
and sets, clears, copies or swaps some bits in that case.

PDL and interference have some differences, though.
PDL deals with arbitrary states, rather than states defined by
a bitfield (i.e.\ an assignment). Consequently,
most presentations of PDL lack assignments that can perform
bit-level operations. This is not a problem for PDL,
since the field has been historically more interested
in the general properties of programs.
However, reasoning about particular programs can be very useful too.
\emph{Dynamic Logic of Propositional Assignments (DL-PA)}~\cite{BalbianiHT13}
operates on a very similar setting to ours,
albeit its assignments can only set or clear bits;
in interference parlance, their redundance notion
is captured by PR~\cite{HeuleKB17}.

Another difference is that programs induced
by existing interference-based proofs are deterministic and complete:
they always yield an output, and the output is unique.
PDL programs are required to be neither.
Furthermore, PDL is typically expressed in a relatively rich
modal propositional language with nesting, disjunction, arbitrary negation
and possibility operators. As we will see,
DSR/WSR proofs seem to implicitly operate over CNF formulas under a
sequence of necessity operators.

In this section, we adapt PDL to our setting based on
conjunctive constraint sets. We will define a notion of programs
closely following \cite{FischerL79,BalbianiHT13};
dealing with the Kleene star operator requires a more detailed
presentation, so we leave it out of the scope of this paper.
We will extend our constraints with a sequence of programs or \emph{context},
which modally correspond to a sequence of necessity operators.
Then we will show that the semantic properties of interference-based
proofs are accurately captured in this framework,
similarly to~\cite{Rebola-Pardo23}.
In fact, the semantics will be captured by entailment,
showing that the reasoning performed by DSR/WSR proofs does not require
interference; rather, the proof rules they use do.

\subsection{Programs as logical relations}
\label{ssc:dynlog:programs}

Let us fix a family of constraints as described in Section~\ref{sec:prelim}.
In what follows, we will refer to these as \emph{static constraints},
and to their associated formulas as \emph{static formulas}.
We define logical relations called \emph{program items}
and \emph{programs} by mutual recursion:
\begin{itemize}
\item (\emph{Assignment}) If $\sigma$ is a finite substitution,
then $\Xang{\sigma}$ is a program item with
$I \otimes J \vDash \Xang{\sigma}$ iff $J = I \circ \sigma$.
\item (\emph{Test}) If $T$ is a static constraint, then $T\test$ is a program item
with $I \otimes J \vDash T$ iff $I \vDash T$ and $I = J$.
\item (\emph{Choice}) If $\varepsilon_1, \varepsilon_2$ are programs,
then $\varepsilon_1 \sqcup \varepsilon_2$ is a program item with
$I \otimes J \vDash \varepsilon_1 \sqcup \varepsilon_2$ iff
$I \otimes J \vDash \varepsilon_1$ or $I \otimes J \vDash \varepsilon_2$.
\item (\emph{Branch}) If $T$ is a static constraint and
$\varepsilon_1, \varepsilon_0$ are programs,
then $\branch{T}{\varepsilon_1 \parallel \varepsilon_0}$ is a program item
with $I \otimes J \vDash \branch{T}{\varepsilon_1 \parallel \varepsilon_0}$
if $I \otimes J \vDash \varepsilon_1$ when $I \vDash T$ and
$I \otimes J \vDash \varepsilon_0$ when $I \not\vDash T$.
\item (\emph{Composition}) If $\varepsilon = \lambda_1\,\dots\,\lambda_n$
is a list of program items, then $\varepsilon$ is a program with
$I \otimes J \vDash \varepsilon$ iff there exist assignments $I_0,\dots,I_n$
with $I_0 = I$, $I_n = J$ and $I_{i-1} \otimes I_i \vDash \lambda_i$
for $0 \leq i < n$.
\end{itemize}
We will use letters $\lambda, \mu$ to denote program items
and $\varepsilon, \delta$ to denote programs.
From an assignments-as-memory-states perspective,
assignments can copy, swap, flip, set and clear predefined bits;
tests are assertions, progressing without change if the test condition holds
and blocking otherwise;
choices non-deterministic execute either of two programs;
and branches execute either of two programs depending on whether
some condition holds.
Programs themselves compose the operations above sequentially,
with the empty program $\emptylist$ behaving as a no-op.
Whenever this is unambiguous, we will identify static constraints $C$
with their empty-context dynamic version $\emptylist.C$.
Observe that under our definitions it is only possible to compose program items,
not programs. However, we can concatenate programs as lists,
which yields the expected semantics of program composition.

Note that our choice of programs is not minimal,
since $\branch{T}{\varepsilon_1 \parallel \varepsilon_0}$ can
be written using the other constructors as
$(T\test\,\varepsilon_1) \sqcup (\overline{T}\test\,\varepsilon_0)$.
The specific form $\branch{T}{\varepsilon_1 \parallel \emptylist}$
is so common we reserve the notation $\branch{T}{\varepsilon_1}$ for it;
this encodes an if-then block, as opposed to the general branch which
encodes an if-then-else block.
Our choice is neither sufficient to express anything we would call a ``program'';
to express loops we would need the Kleene star to construct $\varepsilon^\star$,
which is a program that iteratively applies $\varepsilon$ a non-deterministic
number of times.

\subsection{Dynamic formulas}
\label{ssc:dynlog:formulas}

The core logical expressions in our framework are dynamic constraints
and formulas. Intuitively, a dynamic constraint claims that
a static constraint holds after executing a program,
and dynamic formulas are conjunctions of dynamic constraints.
A \emph{dynamic constraint} is an expression of the form $\varepsilon.C$
where $\varepsilon$ is a program (called the \emph{context})
and $C$ is a static constraint (called the \emph{property}),
with $I \vDash \varepsilon.C$ iff for all assignments $J$ such that
$I \otimes J \vDash \varepsilon$ we have $J \vDash C$.
A \emph{dynamic formula} $\varGamma$ is a finite set of dynamic constraints,
with $I \vDash \varGamma$ iff $I \vDash \varPhi$ for each $\varPhi \in \varGamma$.
We will use letters $\varPhi$, $\varPsi$ to denote dynamic constraints
and $\varGamma, \varDelta, \varTheta$ to denote dynamic formulas.
Given a program $\delta$, a dynamic constraint $\varPhi = \varepsilon.C$
and a dynamic formula $\varGamma$,
we define the dynamic constraint $\delta.\varPhi = (\delta\,\varepsilon).C$
and the dynamic formula $\delta.\varGamma = \Xset{\delta.\varPhi}{\varPhi \in \varGamma}$.
Observe that dynamic formulas are allowed to contain diverse contexts,
so in general we may not be able to write a a dynamic formula
as $\varepsilon.F$.

From a modal logic perspective, each program item in the context
of a dynamic constraint behaves like a necessity operator.
This implies that appending a program context
follows the necessitation and distribution laws~\cite{FischerL79}.
In our setting, this simply means that if an implication is valid,
then it is valid too after under any context.
\begin{theorem}
\label{thm:necessitation}
Let $\varGamma$ and $\varDelta$ be dynamic formulas,
and $\varPi$ be a context.
If $\varGamma \vDash \varDelta$, then $\varPi.\varGamma \vDash \varPi.\varDelta$.
\end{theorem}

Sometimes we will want to restrict a dynamic formula $\varGamma$
to its constraints under a given context $\varepsilon$.
The \emph{contextualization} of $\varGamma$ at $\varepsilon$,
defined by
$\varGamma \at \varepsilon = \Xset{\varPhi}{\varepsilon.\varPhi \in \varGamma}$,
collects all the properties claimed by $\varGamma$ under $\varepsilon$.
note that this includes dynamic constraints whose context
contains additional programs after $\varepsilon$.
Conversely, we might only be interested on the static constraints
in the formula. We define the \emph{static fragment}
of $\varGamma$ as $\varGamma\static = \Xbra{\emptylist.C \in \varGamma}$.

Just like static formulas, dynamic formulas can be reduced
by a substitution $\tau$. We recursively define the \emph{reducts}
of program items, programs and dynamic constraints and formulas
by $\tau$ as follows:
\begin{align*}
\subst{\Xang{\sigma}}{\tau} = {} & \Xang{\tau \circ \sigma} &
\subst{(\varepsilon_1 \sqcup \varepsilon_2)}{\tau} = {} & \subst{\varepsilon_1}{\tau} \sqcup \subst{\varepsilon_2}{\tau} \\
\subst{T\test}{\tau} = {} & \subst{T}{\tau}\test &
\subst{\emptylist}{\tau} = {} & \Xang{\tau} \\
\omit\rlap{$\displaystyle\subst{(\lambda_1\,\lambda_2\,\dots\,\lambda_n)}{\tau} = \subst{\lambda_1}{\tau}\,\lambda_2\,\dots\,\lambda_n$} \\
\omit\rlap{$\displaystyle\subst{(\lambda_1\,\lambda_2\,\dots\,\lambda_n.C)}{\tau} = (\subst{\lambda_1}{\tau}\,\lambda_2\,\dots\,\lambda_n).C$} \\
\subst{(\emptylist.C)}{\tau} = {} & \emptylist.\subst{C}{\tau} &
\subst{\varGamma}{\tau} = {} & \Xset{\subst{\varPhi}{\tau}}{\varPhi \in \varGamma}
\end{align*}

\begin{lemma}
$I \circ \tau \vDash \varGamma$ iff $I \vDash \subst{\varGamma}{\tau}$ holds for any assignment $I$.
\end{lemma}

\subsection{Interference proofs derive dynamic formulas}
\label{ssc:dynlog:interference}

\begin{figure*}
\centering
\begin{tabular}{r@{\hskip2ex}c@{\hskip2ex}l}
\toprule
$\varGamma \Rightarrow \Xang{\sigma}.\varPhi$&$\leadsto$&
    $\varGamma \Rightarrow \subst{\varPhi}{\sigma}$\\[1ex]
$\varGamma \Rightarrow (\varepsilon_1 \sqcup \varepsilon_2).\varPhi$&$\leadsto$&
    $\varGamma \Rightarrow \varepsilon_1.\varPhi$\\[0.3ex] &&
    $\varGamma \Rightarrow \varepsilon_2.\varPhi$\\[1ex]
$\varGamma \Rightarrow T\test.\varPhi$&$\leadsto$&
    $\varGamma \cup \Xbra{T} \Rightarrow \varPhi$\\[1ex]
$\varGamma \Rightarrow \branch{T}{\varepsilon_1 \parallel \varepsilon_0}.\varPhi$&$\leadsto$&
    $\varGamma \cup \Xbra{T} \Rightarrow \varepsilon_1.\varPhi$\\[0.3ex] &&
    $\varGamma \cup \Xbra{\overline{T}} \Rightarrow \varepsilon_0.\varPhi$\\
\bottomrule
\end{tabular}
\qquad
\begin{tabular}{r@{\hskip2ex}c@{\hskip2ex}l}
\toprule
$\varGamma \cup \Xang{\sigma}.\varDelta \Rightarrow \varPhi$&$\leadsto$&
    $\varGamma \cup \subst{\varDelta}{\sigma} \Rightarrow \varPhi$\\[1ex]
$\varGamma \cup (\varepsilon_1 \sqcup \varepsilon_2).\varDelta \Rightarrow \varPhi$&$\leadsto$&
    $\varGamma \cup \varepsilon_1.\varDelta \cup \varepsilon_2.\varDelta \Rightarrow \varPhi$\\[1ex]
$\varGamma \cup T\test.\Delta \Rightarrow \varPhi$&$\leadsto$&
    $\varGamma \cup \varDelta \cup \Xbra{T} \Rightarrow \varPhi$\\[0.3ex] &&
    $\varGamma \cup \Xbra{\overline{T}} \Rightarrow \varPhi$\\[1ex]
$\varGamma \cup \branch{T}{\varepsilon_1 \parallel \varepsilon_0}.\varDelta \Rightarrow \varPhi$&$\leadsto$&
    $\varGamma \cup \varepsilon_1.\varDelta \cup \Xbra{T} \Rightarrow \varPhi$\\[0.3ex] &&
    $\varGamma \cup \varepsilon_0.\varDelta \cup \Xbra{\overline{T}} \Rightarrow \varPhi$\\
\bottomrule
\end{tabular}
\caption{Necessity (left) and possibility (right) dynamic implication rules, where $\varGamma,\varDelta$
are dynamic formulas, $\varPhi$ is a dynamic constraint, $\sigma$ is a finite substitution,
$T$ is a static constraint, and $\varepsilon_0,\varepsilon_1,\varepsilon_2$ are programs.}
\label{fig:rwrules}
\end{figure*}

Let us now go back to DSR and WSR, and construct the program specified
by a derivation by expressing Theorem~\ref{thm:dsr-semantics} within our framework.
To any instruction list $\pi$, we associate the context $\ctx{\pi}$:
\begin{align*}
\ctx{\emptylist} = {} & \emptylist \\
\ctx{\pi\,\del{C}} = \ctx{\pi\,\rup{C}} = {} & \ctx{\pi} \\
\ctx{\pi\,\sr{C}{\sigma}} = \ctx{\pi\,\wsr{C}{\sigma}{G}} = {} &
    \ctx{\pi}\,\branch{\overline{C}}{\Xang{\sigma}}
\end{align*}
The following result shows that \emph{almost} all a refutation is doing
is proving the accumulated formula under a context:
\begin{theorem}
\label{thm:dsr-dynamic}
Let $F$ be a CNF formula and $\pi$ be a derivation from $F$. Then,
$F \vDash \ctx{\pi}.\acc{F, \pi}$.
\end{theorem}
The reason for that ``almost'' is subtle. Assume $\pi$ is a refutation of $F$.
Theorem~\ref{thm:dsr-dynamic} then shows that $F \vDash \ctx{\pi}.\bot$.
This, in principle, is not quite the same as $F \vDash \bot$.
For the program $\ctx{\pi}$, the two are indeed
equivalent\footnote{This can be shown by adapting the proof for the
$\nabla_{\textsc{elim}}$ rule in \cite{Rebola-Pardo23} to the language
of dynamic constraints.}.
However, in general one cannot infer that $F$ is unsatisfiable from
$F \vDash \varPi.\bot$. In fact, $\bot\test.\bot$ is a tautology.

There are three lessons to learn from this.
The first one was already hinted at in \cite{Rebola-Pardo23}:
interference is really an artifact from accumulated formulas.
Theorem~\ref{thm:dsr-dynamic} does not just show satisfiability-preservation,
but entailment. This means that the reasoning by which $\ctx{\pi}.\acc{F, \pi}$
is derived cannot be non-monotonic, since entailment is monotonic.
For example, if two derivations $\pi, \varTheta$ from $F$
have the same context $\varPi = \ctx{\pi} = \ctx{\varTheta}$,
then $F \vDash \varPi.(\acc{F, \pi} \cup \acc{F, \varTheta})$,
regardless that interference-based proof systems
\emph{do not want to allow us} to derive so.

The second lesson is that interference is not even needed
to obtain the full benefits of interference-based proofs.
In fact, this might even be hurtful:
when viewing an accumulated formula from a dynamic perspective,
all dynamic clauses in that formula have the same context,
and some expressivity may be lost to this.
Hence, it seems reasonable to investigate dynamic alternatives
to interference-based proofs.

Finally, the ``almost'' remark above makes apparent
that such dynamic alternatives must provide a way to derive
static conclusions from dynamic premises.
The ultimate reason why interference-based proofs
are so nicely behaved in this regard
is that the programs $\branch{\overline{C}}{\Xang{\sigma}}$
are complete, which is not the case for our programs in general
(e.g.\ $\bot\test$ never progresses).
To derive $\bot$ from $\varepsilon.\bot$
we need to show that \emph{some} execution of $\varepsilon$
progresses. Modally, this involves a possibility operator,
which our dynamic constraints have no way to express.

\section{A dynamic proof system for SAT solving}
\label{sec:proofs}

In this section we provide an interference-free
proof system for dynamic formulas.
To do this, we first define a way to extend conflict detection
to dynamic constraints; in clausal settings,
this can be regarded as ``RUP for dynamic clauses''.
Then we propose proof rules that develop this inference
rule into an interference-free proof system able to prove arbitrary
entailments between dynamic formulas
(as opposed to interference-based proofs,
which can only refute CNF formulas).
Finally, we showcase the expressive power of this
proof system by simulating DSR/WSR proofs through
short, interference-free proofs.

\subsection{Dynamic implication detection}

In this section we extend the conflict detection procedure
over static constraints to conflict detection over dynamic constraints.
The main hurdle we must overcome is that, in general,
we cannot (easily) express the negation of a dynamic constraint
as a dynamic formula, so instead our procedure checks
\emph{dynamic implications} of the form
$\varGamma \Rightarrow \varPhi$ where
$\varGamma$ is a dynamic formula and $\varPhi$ is a dynamic constraint.
A dynamic implication can be seen as the problem of deciding
whether $\varGamma \vDash \varPhi$,
and dynamic implication detection is a sound but incomplete
algorithm for this decision problem.
Our procedure reduces a dynamic implication to a number of
static conflict detection checks;
if they all succeed, then $\varGamma \vDash \varPhi$ holds.

Dynamic implication detection works by iteratively reducing
each implication $\varGamma \Rightarrow \varPhi$ to
several implications $\varGamma_i \Rightarrow \varPhi_i$
for $1 \leq i \leq n$. Reductions,
shown in Figure~\ref{fig:rwrules}, come in
\emph{possibility} and \emph{necessity} variants,
modifying the left- or right-hand side of the implication,
respectively. Albeit it is sound to mix variants,
we will typically only use one set of rules at a time.

Observe that (for a suitably defined program size)
necessity rules always decrease context size
in the right-hand side of an implication,
and possibility rules always decrease
the sum of context sizes in the left-hand side.
Furthermore, some necessity rule can be applied
whenever the right-hand side has a nonempty context,
and some possibility rule can be applied whenever
some constraint in the left-hand side has a nonempty context.
In other words, using both sets of rules we can reduce
a dynamic implication $\varGamma \Rightarrow \varPhi$
to a number of implications of the form
$\emptylist.F_i \Rightarrow \emptylist.C_i$
for static constraints $C_i$, static formulas $F_i$,
and $1 \leq i \leq n$.
We write $\varGamma \dynimp \varPhi$ when
$\conflict{F_i \cup \Xbra{\overline{C_i}}}$
for $1 \leq i \leq n$.
The following result shows dynamic implication detection
extends conflict detection to dynamic constraints:
\begin{theorem}
Consider any dynamic implication rule from Figure~\ref{fig:rwrules}
that reduces $\varGamma \Rightarrow \varPhi$ into
$\varGamma_i \Rightarrow \varPhi_i$ for $1 \leq i \leq n$.
If $\varGamma_i \vDash \varPhi_i$ holds for all $1 \leq i \leq n$,
then $\varGamma \vDash \varPhi$ holds as well.
As a consequence, if $\varGamma \dynimp \varPhi$,
then $\varGamma \vDash \varPhi$.
\end{theorem}

\begin{figure*}
\centering
\begin{tabular}{c@{\hskip5ex}c@{\hskip5ex}c}
$\wsrtrans{F}{\emptylist} = \iqed$&
$\wsrtrans{F}{\rup{C}\,\pi} = \ilem{C}{\iqed}\,\wsrtrans{F \cup \Xbra{C}}{\pi}$&
$\wsrtrans{F}{\del{C}\,\pi} = \wsrtrans{F \setminus \Xbra{C}}{\pi}$\\[1.5ex]
\multicolumn{3}{c}{$\wsrtrans{F}{\sr{C}{\sigma}\,\pi} =
\wsrtrans{F}{\wsr{C}{\sigma}{G}\,\pi} =
\ilem{\varepsilon.\bot}{
    \ilem{\varepsilon.F^\prime}{\iqed}\,
    \ictx{\varepsilon}{\wsrtrans{F^\prime}{\pi}}\,
    \iqed
}\,\ielim\,\iqed$}
\end{tabular}
\caption{A translation from DSR/WSR proofs to proof terms,
where $\varepsilon = \branch{\overline{C}}{\Xang{\sigma}}$ and the formula $F^\prime$ is $F \cup \Xbra{C}$
for the $\sr{C}{\sigma}$ case and $F \setminus G \cup \Xbra{C}$ for the $\wsr{C}{\sigma}{G}$ case.}
\label{fig:wsrtrans}
\end{figure*}

\subsection{Proofs for dynamic formulas}

We now define a proof system in the vein of DSR and WSR for
dynamic formulas. Unlike those systems, our proof system
does not present interference.
In fact, every constraint ever derived is a semantic consequence
of the premises. This has the advantage that proofs can be
nested, since proofs can always be composed.
Proof terms are (possibly empty) lists of instructions,
which define the dynamic constraints we are deriving
and the context under which we are deriving them.

Unlike DSR and WSR, which are only useful to derive unsatisfiability,
these proofs can witness arbitrary entailments
$\varGamma \vDash \varDelta$ between dynamic formulas.
Hence, rather than simply ``checking'' a proof,
we check if a proof $\pi$ witnesses one such entailment.
Below we define the \emph{proof terms} in our system;
an intuitive explanation follows.
For each proof term $\pi$ we give conditions under which
$\pi$ \emph{derives} $\varDelta$ from $\varGamma$,
which we denote $\pproof{\pi}{\varGamma}{\varDelta}$.
\begin{itemize}
\item $\iqed$ is a proof term, with
$\pproof{\iqed}{\varGamma}{\varDelta}$ if for all $\Phi \in \varDelta$
we have $\varGamma\static \dynimp \Phi$.
\item $\ielim\,\pi$ is a proof term if $\pi$ is a proof term, with
$\pproof{\ielim\,\pi}{\varGamma}{\varDelta}$ if
for all $\varPhi \in \varDelta\static$ we have
$\varGamma \dynimp \varPhi$, and
$\pproof{\pi}{\varGamma \cup \varDelta\static}{\varDelta \setminus \varDelta\static}$.
\item $\ilem{\varTheta}{\rho}\,\pi$ is a proof term if $\varTheta$ is a dynamic
formula and $\rho$, $\pi$ are proof terms, with
$\pproof{\ilem{\varTheta}{\rho}\,\pi}{\varGamma}{\varDelta}$ if
$\pproof{\rho}{\varGamma}{\varTheta}$ and
$\pproof{\pi}{\varGamma \cup \varTheta}{\varDelta \setminus \varTheta}$.
\item $\ictx{\varepsilon}{\rho}\,\pi$ is a proof term if $\varepsilon$
is a program and $\rho$, $\pi$ are proof terms, with
$\pproof{\ictx{\varepsilon}{\rho}\,\pi}{\varGamma}{\varDelta}$ if
$\pproof{\rho}{\varGamma \at \varepsilon}{\varDelta \at \varepsilon}$ and
$\pproof{\pi}{\varGamma \cup \varepsilon.(\varDelta \at \varepsilon)}{\varDelta \setminus \varDelta \at \varepsilon}$.
\end{itemize}

Let us consider a proof checker that validates if
$\pproof{\pi}{\varGamma}{\varDelta}$. The proof checker stores a
\emph{database} of derived constraints and a set of \emph{goal}
constraints, initially set to $\varGamma$ and $\varDelta$, respectively.
The checker is also able to perform dynamic implication detection.
Intuitively, $\pi$ is a list of $\welim$, $\wlem$ and $\wctx$
instructions (called \emph{elimination}, \emph{lemma} and
\emph{context}, respectively) terminated by a $\wqed$
(\emph{introduction}) instruction.

$\iqed$ and $\ielim$ are variants of
the same idea: trying to prove that the database implies all goals
by implication detection.
While it is possible to have a rule that uses the whole database to
prove all goals, subsuming $\iqed$ and $\ielim$, this would be
computationally impractical. Figure~\ref{fig:rwrules} reveals that the
number of conflict detection checks grows quickly with the context size
in $\varGamma$ and $\varDelta$. Hence, $\iqed$ only uses static premises,
and $\ielim$ only derives static goals; this means that only
necessity and possibility rules are applied in each respective case.
Operationally, the checker validates the corresponding implication;
in the case of $\ielim$ further goals might still need to be proven,
so $\varDelta\static$ is moved from the goals to the database.

$\ilem{\varTheta}{\rho}$ barely qualifies as a proof rule.
Its only role is to specify intermediate lemmas $\varTheta$ and
to forget temporary constraints used to derive $\varTheta$.
When finding this instruction,
the checker replaces its current goals $\varDelta$ by $\varTheta$,
and validates $\rho$ under the current database $\varGamma$.
Upon success, the original database $\varGamma$ and goals
$\varDelta$ are restored. Since $\varTheta$ has now been derived,
these constraints are added to the database, and removed from the goals
if present.

$\ictx{\varepsilon}{\rho}$ corresponds to Theorem~\ref{thm:necessitation},
deriving every goal under context $\varepsilon$ from the database
premises under $\varepsilon$. To validate this instruction, the checker
sets $\varGamma \at \varepsilon$ as its database and
$\varDelta \at \varepsilon$ as its goals.
After validating $\rho$, the checker restores the database $\varGamma$
and goals $\varDelta$, moving all goals $\varepsilon.\varPhi$
to the database.

We now state the core result of this work: our derivations
form a sound, interference-free proof system for dynamic formulas
with full compositionality.

\begin{theorem}
\begin{enumerate}
\item (Soundness) If $\pproof{\pi}{\varGamma}{\varDelta}$, then $\varGamma \vDash \varDelta$.
\item (Non-interference) If $\varGamma \subseteq \varGamma^\prime$ and $\pproof{\pi}{\varGamma}{\varDelta}$,
then $\pproof{\pi}{\varGamma^\prime}{\varDelta}$.
\item (Compositionality) If $\pproof{\pi_i}{\varGamma_i}{\varDelta_i}$ for $1 \leq i \leq n$,
consider $\pi = \ilem{\varDelta_1}{\pi_1}\,\dots\,\ilem{\varDelta_n}{\pi_n}\,\iqed$.
Then, $\pproof{\pi}{\varGamma_1 \cup \dots \cup \varGamma_n}{\varDelta_1 \cup \dots \cup \varDelta_n}$.
\item (Implicational soundness) For any $\ilem{\varTheta}{\rho}$ instruction in $\pi$
located under nested $\ictx{\varepsilon_i}{\rho_i}$ instructions
for $1 \leq i \leq n$, we have $\varGamma \vDash \varepsilon_1\,\dots\,\varepsilon_n.\varTheta$.
\end{enumerate}
\end{theorem}

\subsection{Short, interference-free proofs for SAT solving}

Just by having a sound proof system we have not accomplished much;
in fact, we are yet to even build a single proof.
Better than that, we now show that these proofs naturally extend DSR/WSR proofs,
themselves a superset of widely used proof systems for SAT solving
such as DRAT and DPR. We do this by translating any DSR/WSR refutation
$\pi$ of $F$ into a proof term $\wsrtrans{F}{\pi}$.
The translation, shown in Figure~\ref{fig:wsrtrans},
is rather straightforward;
the heavy lifting is done by the dynamic implication rules.

\begin{theorem}
\label{thm:dsr-simulation}
Let $\pi$ be a DSR/WSR refutation of a CNF formula $F$.
Then, $\pproof{\wsrtrans{F}{\pi}}{F}{\bot}$.
\end{theorem}
\begin{proof}[Proof sketch]
By induction on $\pi$.
We show the case $\sr{C}{\sigma}\,\pi$;
the case $\wsr{C}{\sigma}{G}\,\pi$ is similar and the others are straightforward.
Let us call $\rho_1 = \ilem{\varepsilon.F^\prime}{\iqed}$,
$\rho_2 = \ictx{\varepsilon}{\wsrtrans{F^\prime}{\pi}\,\iqed}$
and $\rho_3 = \ilem{\varepsilon.\bot}{\rho_1\,\rho_2\,\iqed}$.
Checking $\rho_1$ is equivalent to checking $F \dynimp \varepsilon.D$
for each $D \in F^\prime$. By the rules in Figure~\ref{fig:rwrules},
this elicits the conflict detection (i.e.\ RUP) checks
$\conflict{F \cup \Xbra{C, \overline{D}}}$
and $\conflict{F \cup \Xbra{\overline{C}, \subst{D}{\sigma}}}$.
Since $F^\prime = F \cup \Xbra{C}$, the former is trivial;
and since $C$ is an SR clause over $F$ upon $\sigma$,
the latter holds too.
Furthermore, $\pproof{\rho_2}{\varepsilon.F^\prime}{\varepsilon.\bot}$
by induction hypothesis, so $\pproof{\rho_3}{F}{\varepsilon.\bot}$ holds.
$\ielim$ then requires checking $F \cup \Xbra{\varepsilon.\bot} \dynimp \bot$.
This elicits the trivial conflict detection checks
$\conflict{F \cup \Xbra{\subst{\bot}{\sigma}, \overline{C}, \top}}$
and $\conflict{F \cup \Xbra{\bot, C, \top}}$, which are both trivial.
Hence, $\pproof{\wsrtrans{F}{\sr{C}{\sigma}\,\pi}}{F}{\bot}$ holds.
\end{proof}

This translation slightly increases the size of proofs.
However, we argue this is a matter of formatting.
First, note that $\iqed$ can (and must) only occur at the end of
a proof term, making it essentially an end-of-list marker which can be removed.
Furthermore, missing proofs can be implicitly understood to be $\iqed$.
Second, the formula $F^\prime$ can potentially be enormous.
This can be greatly alleviated by combining DRAT-style cummulative notation
with $\wlem$ instructions to prevent interference-like behavior.
Third, the program $\varepsilon$ is a bit larger than in DPR/DSR proofs;
this can be alleviated introducing program definitions.
Either way, SR instructions are only sparingly used in proofs,
so these details are hardly significant.

Observe that deletions from DSR/WSR proofs are completely
removed in the translation. In fact, our proof terms do not
have a notion of deletion. This is unsurprising:
like other basic interference-free proof systems
(e.g.\ resolution, RUP), deletions are irrelevant for reasoning.
Rather, deletions are merely promises that a constraint
will not be used for the remainder of the proof,
and failing to delete a constraint does not
(or \emph{should not}~\cite{AltmanningerP20})
affect proof correctness.
As a historical note, deletions in DSR-like proof systems
were first introduced in the DRUP proof system~\cite{HeuleHW14}
to improve the efficiency of unit propagation
and to reduce memory footprint.
Deletions only become semantically relevant with the introduction
of the interference-based rule RAT~\cite{JarvisaloHB12}.
Since our proof terms are interference-free,
we can relegate deletions to their original place as
efficiency metadata.

As a capstone, let us demonstrate how this proof system can do
what DSR/WSR cannot: we give a solution to the problem
presented in Section~\ref{ssc:prelim:hurdle}.
The symmetry breakers $B_i$ there are introduced in $G_i$
as SR clauses upon $\tau_i$. Since the rest of the derivation
of $z_i$ is done with a CDCL solver, this is done exclusively
through RUP steps. Following the translation above,
we can construct proof terms
$\pproof{\pi_i}{G_i}{\branch{\overline{B_i}}{\Xang{\tau_i}}.z_i}$.
Now consider the proof term $\rho_i = \pi_i\,\ielim\,\iqed$.
Since $\subst{z_i}{\tau_i} = z_i$, the dynamic implication check
$\branch{\overline{B_i}}{\Xang{\tau_i}}.z_i \Rightarrow z_i$
is rewritten to static conflict checks
$\conflict{\Xbra{z_i, \overline{B_i}, \overline{z_i}}}$ and
$\conflict{\Xbra{z_i, B_i, \overline{z_i}}}$, which obviously succeed.
Then, $\pproof{\ielim}{\branch{\overline{B_i}}{\Xang{\tau_i}}.z_i}{z_i}$ holds,
so $\pproof{\rho_i}{G_i}{z_i}$.
Finally, since $\conflict{\Xbra{z_1, \dots, z_n, C, \top}}$,
the proof term $\rho = \ilem{z_1, \rho_1}\,\dots\,\ilem{z_n, \rho_n}\,\iqed$
satisfies $\pproof{\rho}{G}{\bot}$.

\section{Conclusion}

Dynamic formulas, a logical framework derivative from propositional dynamic logic,
capture the semantics of interference-based proofs.
Theorems~\ref{thm:dsr-semantics}, \ref{thm:dsr-dynamic} and \ref{thm:dsr-simulation}
show this in increasing order of strength:
first we show that the semantics of single inferences are captured,
then we do the same for whole proofs,
and finally we show that the proof structure itself can be simulated
as dynamic proof terms.

The reasoning framework in Section~\ref{sec:proofs} provides an extension
of conflict detection mechanisms like RUP to implications of dynamic formulas.
Since implications (rather than formulas) are detected,
we avoid the need to extend dynamic formulas to deal with negation, disjunction
or possibility operators. In turn, this makes reasoning over incomplete programs
possible.

Building upon this dynamic implication detection method,
we describe a language of proof terms. These proof terms simulate WSR/DSR proofs
in quite an intuitive way. Since proof terms show entailments,
which are monotonic, the usual properties of interference-free proof systems
are recovered, including compositionality.
Furthermore, the framework is carefully tailored to enable efficient checker
implementations based on existing proof checking methods.

\subsection{Future work}

This work glaringly ignores the dominance-based strengthening rule from~\cite{BogaertsGMN23},
an interference-based rule whose proof-theoretic properties have been suggested to be
significantly different from DSR/WSR~\cite{KolodziejczykT24}.
This is an artifact of the need for a reasonable first presentation,
and will be tackled in follow-up work.
The framework presented here was, in fact, inspired by the question:
why does dominance require \emph{two} accumulated formulas?

The omission of dominance is related to another excluded feature, namely the Kleene star program
constructor~\cite{FischerL79}. When viewing interference rules as derivation under programs,
dominance behaves as a while-loop, which in turn can only be expressed using the Kleene star.
Dominance requires two accumulated formulas because they operate under different
contexts. The \emph{core formula} from VeriPB proofs~\cite{BogaertsGMN23} contains
an invariant that holds under a loop program context $(\lambda_1\,\dots\,\lambda_n)^\star$;
the \emph{derived formula} instead holds under each of the $\lambda_1,\dots,\lambda_i$ contexts
for $1 \leq i \leq n$.

Eliminating interference from proofs with dominance is significantly more involved.
For one, dynamic implication for dominance requires some hints (in particular, an inductive invariant)
to work appropriately. For another, proving termination of while-loops
requires the use of a preorder formula over a different set of variables.
Developing this yields a very expressive dynamic logic that can express both regular and concurrent
programs, as well as programs defined by their transition relations \`a la model checking~\cite{BiereCCZ99}.
\bibliography{main}
\end{document}